\documentclass[12pt]{iopart}

\usepackage{iopams}

\begin{document}

\title[Entanglement of spins under  a strong laser influence]{Entanglement of spins under a strong laser influence}

\author{V~Gerdt$^1$, S~Gogilidze$^2$, A~Khvedelidze$^{1,2,3}$,  D~Mladenov$^{4}$
         and V~Sanadze$^2$}

\address{$^1$ Joint Institute for Nuclear Research, Dubna, RU}

\address{$^2$ University of Georgia, Tbilisi, GE }

\address{$^3$  A Razmadze Mathematical Institute, Tbilisi  State University,Tbilisi, GE }

\address{$^4$ Sofia  State University, Sofia, BG }

\eads{\mailto{gerdt@jinr.ru}}

\begin{abstract}
The fate of entanglement of spins for two heavy constituents of a bound state moving in a  strong laser field is analyzed within the semiclassical approach.  The bound state motion as a whole is considered classically beyond the dipole approximation and taking into account the magnetic field effect by using the exact solution to the Newton equation. At the same time the evolution  of constituent spins under the laser influence is studied quantum mechanically. The spin density matrix is  determined as solution to the von Neumann equation with the effective Hamiltonian, describing spin-laser interaction along  the bound state classical trajectory. Based on the solution, the dynamics of concurrence of spins is calculated for the maximally entangled Werner states as well as for an initially uncorrelated state.

\end{abstract}

\pacs{42.50.Ct, 03.65.Ud, 03.67.Bg, 34.80.Qb}

\section{Introduction}

The entanglement of quantum states is a fundamental resource in quantum communication and computation~\cite{GreggJaeger}. In spite of an enormous progress in theoretical and experimental studies, the problem of effective control over entanglement and other quantum characteristics  of multi-particle states yet remains unsolved. Over the last two decades special attention was drown to models based on the usage of a coherent laser radiation. In most cases, for instance in the Cirac-Zoller model of quantum computation with cold trapped ions \cite{CiracZoller},  the laser frequency and polarization play a role of the relevant control parameters.  However, recent impressive success in construction of high-intensity lasers \cite{Mourou} offers an alternative control parameter, \textit{intensity of the coherent electromagnetic radiation}.  Indeed, entering to the regime of a high-intensity, a variety of new physical phenomena has been discovered (see e.g. \cite{SalaminHuHatsagortsyanKeitel}
and references therein).  Very interesting expositions of a strong laser
influence on spins of single particles \cite{HuKeitel} as well on atoms \cite{VrazquezAldanaRoso}
have been predicted.
Evidently, it deserves attention to investigate whether the new  effects due to the laser intensity can be used for the purposes of quantum engineering.

\noindent
Moving towards this direction, it is worth to emphasize
that the  entanglement description in the presence of a strong laser radiation is a highly complex problem owing to the following circumstances:

\begin{enumerate}
\item[\bf{(a)}] The dynamics of a non-relativistic charged particle driven by a
low intensity laser is fully determined by the electric component of electromagnetic field. The electric field dominance together with the dipole approximation provides a consistent solution to the
equation of motion for the classical trajectory of a charged particle \cite{Jackson} in a way not affected its spin  dynamics \cite{ProtopapasKeitelKnight}.
This scheme works perfectly well for various applications of low intensity lasers \cite{DeloneKrainov}.
With a laser intensity large enough the relativistic corrections to the electric charge motion become unavoidable. This requires to abandon the electric dipole approximation and take
into account the magnetic field influence \cite{DruhlMcIver,Reiss1992} on classical trajectory as well as on the particle spin evolution.

\item[\bf{(b)}]
Since dealing with a high-intensity laser the relativistic description
is inevitable, it is necessary to extend the conventional non-relativistic quantum  formalism  for the 
entanglement phenomena to the relativistic case. Today, in spite of undertaken efforts, we are still far from having a complete picture of the relativistic entanglement.

\end{enumerate}

In the subsequent sections, taking into account these observations, we  adopt the conventional semiclassical attitude towards the dynamics of charged particles in an electromagnetic background.
The semiclassical motion of a binary bound state driven by a high-intensity monochromatic and elliptically polarized laser radiation is modelled and studied in the context of dynamics of entanglement. We address the question how the intensity of a laser can affect the correlations between constituent spins. The analysis of the entanglement  in a strong laser background
presented here was developed out of the recent studies~\cite{JamesonKhvedelidze},
\cite{EliashviliGerdtKhvedelidze}.

In the spirit of the WKB approximation the binary composed system motion as a whole is studied classically, assuming that the back reaction of spin of the constituents on the particle dynamics is negligible (cf. this approximation in \cite{RubinowKeller}, \cite{BolteKeppeler}). Our analysis is based on the exact solution to the Hamilton-Jacobi equations for the trajectory of a charged particle interacting with electromagnetic plane wave \cite{JamesonKhvedelidze}. In doing so, the spin evolution will be  treated
quantum mechanically, as required by a spin nature, using the von Neumann equation for the spins density matrix with the leading relativistic corrections included.  Furthermore, the
spin-radiation interaction is encoded in the effective spatially homogeneous Hamiltonian which is determined solely by the bound state classical trajectory (cf. discussions on this approximation in \cite{HuKeitel}).

\section{Semiclassical model for strong laser-bound state interaction}

Consider a massive ($M_B$), electrically charged ($-q_B$) bound state
composed of two charged ($e^{(n)}, e^{(p)}$),  massive  $(m^{(n)}, m^{(p)})$,
spin-1/2 particles interacting with a laser radiation modeled by the monochromatic plane wave propagating along the
$z$-axis
\begin{equation}\label{eq:gpot}
    \boldsymbol{A}(t, \boldsymbol{x}): = a\biggl(\varepsilon\cos(\omega_L\xi)\,,\
    \sqrt{1-\varepsilon^2}\sin(\omega_L\xi)\,,\ 0
    \biggl)\,, \qquad \xi= t-\frac{z}{c}\,.
\end{equation}
In (\ref{eq:gpot}) the parameter $\varepsilon\in [0,1]\, $ denotes  the
light polarisation,  $\omega_L$ is the wave frequency. The constant ${a}$ determines
the dimensionless laser field strength parameter~\cite{Sengupta,SarachikSchappert}
\[
\eta^2=\frac{q_B^2\,a^2}{M_B^2\,c^4}
\]
setting the scale for the intensity of a laser-bound state  interaction.

The semiclassical picture is mathematically formulated as follows.
Let us divide all configuration variables into  three parts:
\textit{the center-mass coordinate 
\[
\boldsymbol{R}=(m^{(n)}\boldsymbol{r}_n+
m^{(p)}\boldsymbol{r}_p)/{M_B}\,,
\]
the relative coordinate between  constituents,
$\boldsymbol{r}=
\boldsymbol{r}_n-\boldsymbol{r}_p\,,$ and their spin
variables}. Correspondingly, the Hilbert space $\mathcal{H}$
is decomposed as
\[
\mathcal{H}
= \mathcal{H}_{\mathrm{CM}}\otimes\mathcal{H}_{\mathrm{RM}}
\otimes\mathcal{H}_{\mathrm{SPIN}}\,.
\]

The dynamics on  $\mathcal{H}$ is assumed to be driven by the following  interactions:
\begin{itemize}
\item The electric charge of the bound state has a point-like distribution,
piked at the position of its center of mass  $\boldsymbol{R}$ with
a  ``point-like'' part  of the laser-charge interaction $V_{CL}$
described by the conventional radiation scattering of the electric charge $-q_B$, moving with the velocity
$\boldsymbol{\upsilon}_R= {d}\boldsymbol{R}/{{d} t}\,$
\[
    V_{CL}:=
    \frac{q_B}{c}\,\boldsymbol{\upsilon}_{R}\,\cdot\boldsymbol{A}(t,\,
    \boldsymbol{R})\,.
\]

\item  The degrees of freedom of  the  constituents evolve in time and 
interact with each other ($V_B$) as well as with a laser radiation ($V_{SL}\,$)
\[
    V_B = V_0(r)+ V_{SS}(r)\,, \qquad
    V_{SS}(r) := V_S (r)\,\boldsymbol{S}\otimes\boldsymbol{S}\,,
\]
where  $V_0(r)$ and $V_S(r)$ are scalar functions of the relative distance $r=
|\boldsymbol{r}_n-\boldsymbol{r}_p|\,$  between constituents, and the Pauli matrices  $\boldsymbol{\sigma}=(\sigma_1,\sigma_2,\sigma_3)$ are used
to describe the spin of constituents. 
$\boldsymbol{S}=\frac{\hbar}{2}\,\boldsymbol{\sigma}$.
The spin-laser coupling $V_{SL}$ is determined by the magnetic moments of constituents in the relativistically modified Larmor form
\[
V_{SL}:= - \boldsymbol{\Omega}^{(n)}(t,\boldsymbol{r}_n) \cdot
\boldsymbol{S}^{(n)}- \boldsymbol{\Omega}^{(p)}(t,\boldsymbol{r}_p)
\cdot \boldsymbol{S}^{(p)}\,,
\]
where vector $\boldsymbol{\Omega}^{(i)}\,$ reads
\begin{equation}\label{eq:eff.mag}
  \boldsymbol{\Omega}^{(i)} := \frac{e^{(i)}\,\mathrm{g}^{(i)}}{2\,m^{(i)}c}
\left(\boldsymbol{B}- \frac{1}{c}\, \left[\boldsymbol{v}^{(i)}\times
\boldsymbol{E}\right]\right)\, + \, {\frac{1}{2\,c^2}\,
\left[\boldsymbol{v}^{(i)}\times \boldsymbol{a}^{(i)}\right]}\,.
\end{equation}
$\boldsymbol{E}$ and $\boldsymbol{B}$ in (\ref{eq:eff.mag}) are  respectively the electric and magnetic components of a laser field evaluated along the trajectory of  $i$-th particle (with the gyromagnetic ratio $g^{(i)}$)
moving with the velocity $ \boldsymbol{v}^{(i)}$ and acceleration $\, \boldsymbol{a}^{(i)}$
seen in the LAB frame. The term in parenthesis is the magnetic
field in the instantaneous rest frame of a charged particle (Galilei boosted) while the last contribution in (\ref{eq:eff.mag}) is the leading relativistic Thomas precession correction  \cite{Thomas}
due to the non-vanishing curvature of the particle trajectory, cf. e.g. \cite{Jackson}.
\end{itemize}

\noindent
Gathering all the above together, the evolution of a bound state
travelling in the laser background is governed by the total Hamiltonian
\[
    H= H_0+V_{SS}+V_{CL}+V_{SL}\,,
\]
where $H_0$ is the Hamiltonian of free spinless constituents.

\section{The terse summary of computation}

\subsection{Evolution of center-mass motion of bound state}

In the leading semiclassical approximation  the
contribution to the phase of wave function that comes from the laser-spin interaction term $V_{SL}$  is negligibly small. This term will come into play later on, when we turn to
a study of dynamics of spin degrees.
Therefore  the density matrix of our system admits the \textit{charge $\& $ spin decomposition}
\begin{equation}\label{eq:spin-charge.decomp}
\rho=\sum_{\alpha=\pm}\ c_{\alpha}|\psi_\alpha\rangle\otimes \varrho_{\alpha}\,,
\end{equation}
where two states $|\psi_\pm\rangle$ are  linearly independent WKB
solutions to the Schr\"{o}dinger equation with the Hamiltonian $H_0 + V_{SS}+V_{CL}\,$ and $\varrho_\alpha$
is the density matrix of constituent spins.
This Hamiltonian admits separation of the relative and absolute motion, and for the last one  can use
{\it the exact solution}  \cite{JamesonKhvedelidze} to the analogous Hamilton-Jacobi
problem for a point-like charged particle. 
According to \cite{JamesonKhvedelidze}, the Hamilton-Jacobi generating function for a spinless particle travelling in an arbitrary plane wave background of the form  $A_\mu:=(0,\boldsymbol{A}_\bot(\xi), 0 )$ reads 
\begin{equation}\label{eq:solF}
 \mathcal {F}(\xi\,, \mathbf{\Pi}) = - c(mc-\Pi_z)\,\xi +
   c\int_0^\xi \mathrm{d}u\,
 \sqrt{(mc-\Pi_z)^2+ W(u, \mathbf{\Pi}_\bot)}\,,
\end{equation}
where
\[
    W(\xi, \mathbf{\Pi}_\bot):=-\frac{e^2}{c^2}\,\boldsymbol{A}^2_\bot
    +2\,\frac{e}{c}\,\boldsymbol{A}_\bot\cdot\mathbf{\Pi}_\bot\,.
\]
The constants $\Pi_z$ and $\Pi_\bot$ are determined from the
initial  value of the particle velocity.  With the aid of (\ref{eq:solF}) the standard calculations
give the leading semiclassical wave function
\[
    \langle\boldsymbol{x}, t|\psi_+\rangle=
    \frac{1}{\sqrt{|\partial \mathcal{F} /\partial \xi|}}\,
e^{\frac{i}{\hbar}\boldsymbol{\Pi}_\bot\cdot\boldsymbol{x}_\bot}\,
e^{\frac{i}{\hbar}\mathcal{E}t}\,
 \exp{\frac{i}{\hbar}
{\mathcal{F} (\xi, \boldsymbol{\Pi}_\bot)}}\,.
\]
To make formulas more compact let impose the initial condition
on the classical trajectory  $\boldsymbol{R}(t=0)=0\,$  and fix the
frame, where the time average value of the component of particle velocity
orthogonal to the electromagnetic wave propagation direction
vanishes, \(\, \langle\langle\boldsymbol{\upsilon}_\bot
\rangle\rangle=0\,. \) From the generating function (\ref{eq:solF}) it follows that
 the bound state center of mass moves  along  the trajectory:
\begin{eqnarray}
R_x(t) &=&
  -\frac{c}{\omega_L}\,\sqrt{\frac{\varepsilon^2}{1-2\,\varepsilon^2}}
  \,\arcsin \left[\mu\,
  \mathrm{sn}\big(u,\, \mu\big)\right]\,, \label{trajectoryLAB-x} \\
 R_y(t)&=& \frac{c}{\omega_L }\,\sqrt{\frac{1-\varepsilon^2}{1-2\,\varepsilon^2}}
  \,\ln \left[
\frac{\mu\,\mathrm{cn}\big(u,\, \mu\big)
+\mathrm{dn}\big(u,\, \mu\big)}{1+\mu}\right]\,,\label{trajectoryLAB-y} \\
R_x(t) &=&  c t - \frac{c}{\omega_L}\,\mathrm{am}\big(u,\, \mu\big)\,. \label{trajectoryLAB-z}
\end{eqnarray}
The trajectory (\ref{trajectoryLAB-x})-(\ref{trajectoryLAB-z}) is expressed in
terms of the Jacobian elliptic functions $\mathrm{sn}(u,\mu), \
\mathrm{cn}(u,\mu), \ \mathrm{dn}(u,\mu) $ and the amplitude function
$\mathrm{am}(u,\mu)$ \cite{WhittackerWatson}.
 The argument $u:= \omega_{L}^\prime t$ of these functions
is the laboratory frame time $t$ scaled by the 
 laser frequency $\omega_L^\prime = \gamma_z\,\omega_L$ non-relativistically  Doppler shifted
  by  $ \gamma_z=1-{\upsilon_z(0)}/{c}$.
The modulus $\mu$ is determined by the laser and the particle characteristics
\[
\gamma_z^2\,\mu^2=(1-2\,\varepsilon^2)\,\eta^2\,.
  \]

In  (\ref{trajectoryLAB-x})-(\ref{trajectoryLAB-z}) the modulus belongs  to the fundamental domain $ 0 < \mu^2 < 1$.  The solution  outside this interval
 can be reconstructed from it by using the modular properties of the Jacobian functions.
For the corresponding details  we refer again to  \cite{JamesonKhvedelidze}.

\subsection{The evolution of spin degrees }

The semiclassical calculations imply that  the spin  density matrix
$\varrho$ in the decomposition
(\ref{eq:spin-charge.decomp}) satisfy the \textit{spin evolution
equation} written  in the form of von Neumann equation
\begin{equation}\label{eq:Neumann}
   \dot\varrho(t)=-\frac{i}{\hbar}\,[H_S(t),\, \varrho(t)]\,.
\end{equation}
The effective spin Hamiltonian $H_S$ is defined as
the projection of the Hamiltonian $V_{SS} + V_{SL}$ to
the classical trajectory of the constituent particles:
\begin{equation}\label{eq:effHam}
    H_S(t) = V_{SS} +
    V_{SL}\biggl|_{\mbox{Particles~classical~trajectory}}\,.
\end{equation}
To evaluate  (\ref{eq:effHam}) we follow the spirit of  the Born-Oppenheimer approximation
\cite{Born-Oppenheimer}.
Namely,  we ``freeze'' the relative motion of
constituents inside the bound state, i.e. approximate their
relative trajectory by the mean value $<r(t)>= 0$ and
neglect all contributions of order $\upsilon_r/c$,
where $\upsilon_r$ is the relative velocity of constituents.
A straightforward evaluation of the effective laser-spin Hamiltonian (\ref{eq:effHam}) gives
\begin{equation}\label{eq:2spinHam}
    H_S = - \boldsymbol{\mathfrak{B}}^{(n)}(t)\cdot
    \boldsymbol{S}\otimes I -  I\otimes\boldsymbol{S}\cdot
    \boldsymbol{\mathfrak{B}}^{(p)}(t) + H_I\,,
\end{equation}
where $\mathfrak{B}^{(i)}(t)\,,\,
i=(n,p)$ for $\,\widetilde{\mathrm{g}}^{(i)}=\left(e^{(i)}/m^{(i)}\right)\left(
M_B/q_B\right)\,{\mathrm{g}}^{(i)}$ read
\begin{eqnarray*}
\mathfrak{B}^{(i)}_x(t)
&=&\eta\frac{\omega_L^\prime}{2}\sqrt{1-\varepsilon^2}
   \left[(\widetilde{\mathrm{g}}^{(i)}+1)\mathrm{dn}(u,\,\mu)-\gamma_z\right]
\mathrm{cn}(u\,,\mu)\,,
   \\
\mathfrak{B}^{(i)}_y(t)&=&
\eta\frac{\omega_L^\prime}{2}\,\varepsilon
   \left[(\widetilde{\mathrm{g}}^{(i)}+1)\mathrm{dn}(u\,\mu)-\gamma_z(1-\mu^2)\right]
\mathrm{sn}(u\,,\mu)\,, \\
\mathfrak{B}^{(i)}_z(t) &=&
   -\eta^2\frac{\omega_L}{2}\,\varepsilon\sqrt{1-\varepsilon^2}
   \left[{\widetilde{\mathrm{g}}^{(i)}}- {\gamma_z}\mathrm{dn}(u\,,\mu)\right]\,.
\end{eqnarray*}

Similarly, the spin-spin interaction term $H_I$ in (\ref{eq:2spinHam})
originates from $V_{SS}\,$ under the same static
approximation for the spatial relative degrees of freedom:
\begin{equation} \label{eq:InterHam}
{\hbar} H_I={g}\,\boldsymbol{S}\otimes\boldsymbol{S}\,.
\end{equation}
The constant $g$ in (\ref{eq:InterHam}) is determined by the 
spin-spin potential evaluated at the mean value of the relative
distance between the constituents, $g: = \hbar\,V_S(0)\,.$

\section{Dynamics of entanglement}

Now one can analyze  the dynamics of entanglement under the environment coupling
\cite{MinAndCarvalhKusbBuchlei} realized in our model by a
background laser radiation. We postpone for a future analysis  a generic
case  and consider the dynamics of density matrices of a special
type only.

\subsection{Werner states.}

Consider first  a family of entangled mixed states,
the so-called Werner states \cite{Werner},  characterized
by a single real parameter $p$ that measures the overlap of a given Werner
state with the maximally entangled pure Bell state
\begin{equation}\label{eq:Werner}
    \varrho_{\scriptsize{W}}:=\frac{1}{4}
    \left(\mathrm{I}+ p\,\boldsymbol{\sigma}\otimes\boldsymbol{\sigma}
\right)\,.
\end{equation}
For $\frac{1}{3}< p \leq 1$  the  density matrix (\ref{eq:Werner}) describes the mixed entangled state.

To find the fate of the entanglement of the initial Werner state
we use the expression for the evolution operator $U(t)=X(t)W(t)$  given in Appendix.
Since the entanglement properties are invariant under the
local unitary transformation of the form $W(t)=U_{(n)}(t)\otimes U_{(p)}(t)$, only the action of the operator
$X(t)$ may affect the entanglement. With this observation one can easily
evaluate the leading, in a laser intensity, change of the
density matrix
\[
\delta_t\varrho_{W} =\frac{i}{\hbar}[V_I, \varrho_W ]\,.
\]
This gives the expression
\[
\delta_t\varrho_{W}=-\frac{1}{2}g
\eta\,p\Delta
\sin(\omega_L t)\,\left[\cos(4gt)\,\sigma_{[30]}
    +\frac{3}{4}\sin(4gt)\,
    \sigma_{[12]}\right]\,,
\]
where
\[
\Delta=\widetilde{\mathrm{g}}^{(n)}-\widetilde{\mathrm{g}}^{(p)}\,,\quad  \sigma_{[\mu\nu]}:=\sigma_\mu\otimes\sigma_\nu-\sigma_\nu\otimes\sigma_\mu\ (\mu,\nu=0,1,2,3)\,,
 \]
and $\sigma_0$ is the unit $2\times 2$ matrix. As a result, one can obtain that in the leading order in a laser intensity
the concurrence is stable under the influence of laser background:
\[
C(  \varrho_{W} )=  \max \left(0, \frac{3p-1}{2} \right)\,.
\]

\subsection{Initially uncorrelated spins}

The same strategy can be applied to the initially  uncorrelated spin state
\[
    \varrho_0 = \frac{1}{4}\,\left(\mathrm{I}+\alpha\,
\frac{1}{2}\left(\sigma_{03}+\sigma_{30}\right)+
\beta\,\frac{1}{2}\left(\sigma_{03}-\sigma_{30}\right)
\right)\,.
\]
Our calculations show  that the concurrence, in the leading order in {$ \eta$}, is
\begin{equation}\label{eq:concprod}
C(  \varrho_0 ) =  max \left(  0, 4\eta |\beta g\Delta Q(t)| -  \sqrt{1-\alpha^2}\right)\,,
\end{equation}
where
$$Q (t) = \displaystyle \frac{1}{\omega_L+4g}\sin^2(\omega_L/2+2g)t+  \frac{1}{\omega_L-4g}\sin^2(\omega_L/2-2g)t\,.$$

This example demonstrates  the possibility of formation of the entanglement solely due to the effects of a laser field intensity. Note that, remaining within the dipole-approximation and ignoring   the intensity influence on the spin dynamics,   we will get  the vanishing  concurrence  rather than  the above  derived result  (\ref{eq:concprod}).

\section{Concluding remarks}

In the present note to understand the dynamics of entanglement under 
strong field influence we formulated the model for the bound state composed of two heavy charged
spin-1/2 particles traveling in the laser field. The relative motion of constitutes was treated within
the Born-Oppenheimer method~\cite{Born-Oppenheimer}
and the semiclassical approximation has been used to find the
evolution operator. Our studies show the following:
\begin{itemize}
\item The entanglement between constituents
with the different gyromagnetic ratios evolves in a manner
strongly depending on the intensity of a laser beam
as well as on the coupling between spins.
\item There are cases when the entanglement properties reveal stability
in the leading order. This happens particularly for the Werner states of  the spin density
marix.
\item There is a possibility to attain the entanglement manipulating with  the
laser intensity. Particularly, in our model uncorrelated initial state evolving in a laser field  squires nontrivial
  concurrence merely due to the strong laser intensity.

\end{itemize}
The last observation may have an impact on very interesting and promising studies of 
the nonlinear effects in spin-laser interactions.  It may open an alternative way for the manipulation with entanglement of spins  using  a laser intensity as a control parameter.

It is worth to note that  the methods suggested in the article are adequate and well adapted to  the description of  processes in transition from the non-relativistic to relativistic regime only.  For completion of studies the fully relativistic treatment  is necessary.  Today its elaboration   remains an open and  intrigued research area.

\ack

This work was partially supported by the grant No.10-01-00200 from Russian Foundation for Basic Research  (the contribution of V.G. and A.Kh.), by the Collaborative grant ``Bulgaria-JINR'' (the contribution of A.Kh. and D.M.), by the ``Holubei-Meshcheryakov'' program (the contribution of V.G.), by the grant of the University of Georgia (the contribution of S.G. and V.S.) and by the grant 148/2012 from the Sofia University Research Fund (the contribution of D.M.).
\appendix
\setcounter{section}{1}

\section*{Appendix}

Consider  (\ref{eq:Neumann})  for spins evolving in the  linearly polarized plane wave, $(\varepsilon=0)\,.$
The spin  density matrix $\varrho \,,$ obeying  a certain  initial condition at $t=0\,,$
is given by the unitary evolution  operator $U$
$$\varrho(t)  = U(t) \varrho(0)\, U^+(t)\,.$$
It is convenient to pass  to the ``interaction picture'' $U(t)=W(t)X(t)$,
i.e.  factor  out from $U$  the operator  $W(t):=U^{(n)}(t)\otimes U^{(p)}(t)$  that describes the dynamics  of  spins,$ (n) $ and $(p)$,  in a laser background, but not interacting with each other while precessing.
The unknown  operator $X(t)$ is subject to the equation
\begin{equation}\label{eq:evoleq}
    \dot X(t) = - \frac{i}{\hbar}\, H_I^\prime(t)\, X(t)
\end{equation}
with the  Hamiltonian (\ref{eq:InterHam}) written in  the ``interaction picture''
\begin{equation}\label{eq:hamint}
H_I^\prime(t):= W^+ H_I W ={g}\hbar\,\left(
\cos\vartheta_-\,\boldsymbol{\sigma}\otimes\boldsymbol{\sigma}
+\sin\vartheta_-\,\sigma_{[12]}\right)\,,
\end{equation}
where
\[
\vartheta_-(t):=\vartheta^{(n)}-\vartheta^{(p)}=\frac{1}{2}\,
\eta\,\Delta\,
\mathrm{sn}(u, \,\mu)\,.
\]
To obtain  (\ref{eq:hamint})  the expression
 $U^{(i)}(t) =  \exp\left( \frac{i}{2}\, \vartheta^{(i)}(t)\,\sigma_1\right)$  with
\[
2 \vartheta^{(i)}(t)=
{\eta}\,\left(\widetilde{\mathrm{g}}^{(i)}+1\right)\,\mathrm{sn}(u,\mu)-\arcsin\left(\mu\,\mathrm{sn}(u, \mu)\right)
\]
has been used.
Note, that $\vartheta^{(i)}(t)$  depends nonlinearly on the laser intensity $\eta$,  and only for small intensities $\eta \ll 1$ reduces to the well-known expression for the angle characterizing the non-relativistic precession
\[
    2\vartheta_{NR} =\eta\,\widetilde{\mathrm{g}}^{(n)}\,\sin(\omega_Lt)\,.
\]

The solution to (\ref{eq:evoleq}) is expressed in terms of the time-ordered exponent
$$ X(t)=\displaystyle e^{ ig\psi(t)
\,\boldsymbol{\sigma}\otimes\boldsymbol{\sigma}}\,{T}\left(
\exp \frac{i}{\hbar}\,\int_0^tV_I(\tau)d\tau\right).
$$
Here
\[
\psi(t)= \int_0^t dt \cos\vartheta_-(t)\,
\]
and
\[
    V_I(t):= g\hbar\,\sin\vartheta_-\left[\cos(4g\psi(t))\,\sigma_{[12]}
    +\frac{3}{4}\sin(4g\psi(t))\,
    \sigma_{[30]}\right].
\]
For the first factor in $X(t)$ one can use the remarkable Eulerian representation
\[
\displaystyle e^{\displaystyle i\psi\,
\,\boldsymbol{\sigma}\otimes\boldsymbol{\sigma}}= \frac{1}{2}\,
 e^{\displaystyle i\psi}+
 \frac{1}{2}\,e^{\displaystyle -i\psi}\,
 \bigg[ \cos 2 \psi +
 i\,\boldsymbol{\sigma}\otimes\boldsymbol{\sigma}\,\sin 2
 \psi\bigg]\,.
\]


\end{document}